\documentclass[submission,copyright]{eptcs}

\providecommand{\event}{HCVS 2016} % Name of the event you are submitting to
\usepackage{listings} % Source Code listings.
\usepackage{graphicx} % Allows inclusion of JPEGs.
\usepackage{verbatim} % Allows begin{comment}
\usepackage{ifthen}
\usepackage{xspace}
\usepackage{amssymb, amsmath, stmaryrd, url}
\usepackage{tikz}
\usetikzlibrary{arrows}
\usepackage{pstricks,pst-node,pst-tree}
\usepackage{pifont}
\usepackage{varwidth}
\usepackage{hhline}
\usepackage{multirow}
\usepackage{multicol}
\usepackage{rotating}
\usepackage{latexsym}
\usepackage{etex}
\usepackage{todonotes}
\usepackage{wrapfig}
\usepackage{enumitem}
\usepackage{color}

\newtheorem{definition}{Definition}
\newtheorem{theorem}{Theorem}

\newcommand{\ignore}[1]{}

\title{Removing Unnecessary Variables from \\Horn {{Clause}} Verification Conditions}
\author{
  Emanuele De Angelis\thanks{Research associate at CNR-IASI, Rome, Italy.}~ and Fabio Fioravanti$^*$
  \institute{DEC, University ``G.~d'Annunzio" of Chieti-Pescara, Italy}
  \email{\{emanuele.deangelis, fabio.fioravanti\}@unich.it}
  \and Alberto Pettorossi$^*$
  \institute{DICII, University of Rome Tor Vergata, Italy}
  \email{adp@iasi.cnr.it}
  \and Maurizio Proietti
  \institute{CNR-IASI, Rome, Italy.}
  \email{proietti@iasi.cnr.it}
}

\def\titlerunning{Removing Unnecessary Variables from Horn Verification Conditions}
\def\authorrunning{E. De Angelis, F. Fioravanti, A. Pettorossi \& M. Proietti}

\begin{document}
\maketitle

\begin{abstract}
Verification conditions (VCs) are logical formulas whose satisfiability 
guarantees program correctness.
We consider VCs in the form of constrained Horn clauses (CHC)
which are automatically generated from the encoding 
of (an interpreter of) the operational semantics of the programming language. 
VCs are derived through program specialization 
based on the unfold/fold transformation rules and,
as it often happens when specializing interpreters, 
they contain {\em unnecessary} variables, that is, variables which are not 
required for the correctness proofs of the programs under verification.
In this paper we adapt to the CHC setting some of the techniques that were developed 
for removing unnecessary variables from logic programs,
and we show that, in some cases, the application of these techniques
increases the effectiveness of Horn clause solvers 
when proving program correctness.
\end{abstract}

\section{Introduction}

Correctness of an imperative program $P$ can be verified 
by: first, (i)~generating \textit{verification conditions} (VCs, for short)
for the program $P$ and the considered property,
and then, (ii) using SMT solvers for checking the satisfiability of the VCs.

In this paper we consider VCs which are automatically derived
by applying program specialization to a 
constrained Horn clause encoding 
of the operational semantics of the programming language.
(In this paper we will use the notions of \textit{Constrained Horn Clauses} (CHC) 
and \textit{constraint logic programs} (CLP) interchangeably.) 
Program specialization is based on the application of 
semantics preserving unfold/fold transformation rules,
guided by a strategy, called {the} {\it VCG strategy}, which has been specifically 
designed for VCs generation {(see \cite{De&15b} for a detailed presentation)}.
{Other} notable applications of CLP program specialization 
to the analysis of imperative or object-oriented programs 
can be found in~\cite{Al&07,Pe&98}.

Given an imperative program $P$ and a safety property, 
we introduce a CLP program~$I$, 
which defines the nullary predicate \texttt{unsafe} such 
that~$P$ is safe if and only if the atom 
\texttt{unsafe} is not derivable from~$I$
or, equivalently, \texttt{unsafe} does not belong to the  \textit{least model} of~$I$, denoted $\mathcal M(I)$.

The \textrm{VCG} strategy works by
performing the so-called \textit{removal of the interpreter},
that is, it removes  
the level of interpretation which is present
in the initial CLP program $I$, where commands are encoded as
CLP clauses and there are 
references to the operational semantics of the imperative 
programming language. 
The output of the \textrm{VCG} strategy is  a program $I_{\it sp}$
such that $\texttt{unsafe} \in \mathcal M(I)$ iff 
\mbox{$\texttt{unsafe}\in\mathcal M(I_{\textit{sp}})$}.
Moreover, due to the absence of the interpretative level, 
the test of whether or not
$\texttt{unsafe}$ belongs to~$\mathcal M(I_{\textit{sp}})$ 
is often easier than the test of whether or not
$\texttt{unsafe}$ belongs to~$\mathcal M(I)$.

\smallskip

{The specialization-based approach} for generating VCs 
is parametric with respect to: (i)~the imperative program $P$, 
(ii)~the operational semantics of the imperative language in which the program $P$
is written, 
(iii)~the property to be proved,
and (iv)~the logic used for specifying the property of interest
(in this case, the reachability of an unsafe state).

One of the most significant advantages of this approach
is that it enables the design of widely applicable 
VC generators for programs written in different 
programming languages, and for different operational semantics 
of languages with the same syntax, by making
small modifications only~\cite{De&15b}.

\section{Removing Unnecessary Variables}
\label{sec:auxtransf}

It is well known that program specialization and transformation techniques
often produce clauses with more arguments than
those that are actually needed~\cite{LeS96,PrP95a, GaK14a}.
Thus, it is not surprising to observe that such a side-effect  
also occurs when generating VCs  via program specialization.
Indeed, it is often the case 
that some of the variables occurring in the CLP program
$I_{\it sp}$, which is generated by the VCG strategy,
are not actually needed to check whether or not $\texttt{unsafe} \in \mathcal M(I_{\it sp})$.
Avoiding those unnecessary variables, and thus deriving predicates with
smaller arity, can increase the effectiveness 
and the efficiency of applying Horn clause solvers, and proving program correctness.

Now we present two transformation techniques which allow us to reduce the number of arguments of the predicates used in the VCs.
These techniques extend to the case of CLP programs similar techniques that have been developed for logic programs~\cite{LeS96,PrP95a}.
The first technique is 
a transformation strategy, called the
\textit{Non-Linking variable Removal strategy} (or the \textit{NLR strategy}, 
for short) 
that removes variables 
occurring as arguments of an atom in the body of a clause, but 
that do not occur elsewhere in the clause.
The second technique, called the \textit{constrained~FAR algorithm}  
(or the \textit{cFAR algorithm}, for short), 
is a generalization of a liveness analysis, and removes arguments that are not 
actually used during program execution.

\vspace{-2mm}

\paragraph{1. Non-Linking variable Removal Strategy{\rm{.}}} 
First, we consider the NLR strategy whose objective 
is to remove the \textit{non-linking variables}. They are defined as follows. 

\begin{definition}[Linking variables~{\rm{\cite{PrP95a}}}]{\rm
Let $C$ be the clause $\texttt{H\,:-\,c,\,L,\,B,\,R}$, where: \texttt{c} is a constraint,
\texttt{L} and~\texttt{R} are (possibly empty) conjunctions of atoms, 
and \texttt{B} is an atom. 
The set of \textit{linking variables} of \texttt{B} in~$C$, denoted by 
$\textit{linkvars}(\texttt{B},C)$, 
is $\textit{vars}(\texttt{B})\cap\textit{vars}(\{\texttt{H,c,L,R}\})$.
The set of \textit{non-linking variables} of \texttt{B} in~$C$ is 
$\textit{vars}(\texttt{B})-\textit{linkvars}(\texttt{B},C)$.}
\end{definition}

Before presenting the NLR strategy, we see it in action in an example.
Let us consider the C program~$P$ in Figure~\ref{fromCtoVCs}. We
want to verify the Hoare triple $\{{\tt x}\geq0\}\ P\ \{{\tt y}\leq0\}$.
By applying the VCG strategy, we get the set of 
clauses $P1$ in Figure~\ref{fromCtoVCs}, where \texttt{unsafe} holds iff
the Hoare triple is not valid.
In~$P1$ 
the non-linking variables have been underlined.
Then, by applying the NLR strategy, we get 
the set of clauses~$P2$ 
without non-linking variables (see Figure~\ref{VCs-by-NLR-and-cFAR}).
$P1$ and $P2$ are  
equivalent with respect to the query
\texttt{unsafe}, in the sense that 
$\texttt{unsafe}\! \in\! \mathcal M(P1)$ 
iff $\texttt{unsafe}\!\in\! \mathcal M(P2)$.

\begin{figure}
	\begin{minipage}{1.0\linewidth}
		\vspace{2mm}
		{\small
			\begin{multicols}{2}
				\begin{enumerate}[topsep=-1pt,parsep=-1pt,partopsep=2pt,
					itemsep=1pt,labelsep*=0pt,leftmargin=15pt,label=\arabic*]	
					\item[]\label{auxex9} {\tt int x,y;}	
					\item[]\label{auxex8} {\tt void main() \{ }
					\item[]{\tt	~~int z=x+1; }	
					\item[]\label{auxex7} {\tt~~while(z<=9) }	
					\item[]{\tt	~~~~~~z=z+1; }
					\item[] {\tt  ~~y=z; }	
					\item[]\label{auxex6} {\tt \} }	
					\item[]
					\vspace{-1mm}
					\item[]{\normalsize{The C program $P$}}
				\end{enumerate}	
				\columnbreak				
				\begin{enumerate}[parsep=-1pt,partopsep=-2pt,
					itemsep=1pt,labelsep*=5pt,leftmargin=-20pt]
					\item\label{auxex4}{\tt	unsafe:- X1>=0, Y2=<0, newp1(X1,\underline{Y1},\underline{X2},Y2). }
					\item\label{auxex3}{\tt	newp1(X1,Y1,X2,Z2):- Z1=X1+1, }
					\item[]{\tt	~~~~newp2(X1,Y1,Z1,X2,\underline{Y2},Z2). }
					\item\label{auxex2}{\tt	newp2(X1,Y1,Z1,X2,Y2,Z2):- Z1=<9,~Z3=Z1+1,} 	
					\item[]{\tt	~~~~newp2(X1,Y1,Z3,X2,Y2,Z2). }
					\item\label{auxex1}{\tt	newp2(X1,Y1,Z1,X1,Y1,Z1):- Z1>=10. }
					\item[]
					\item[]
					\vspace{-1mm}
					\item[]{\normalsize{\hspace{-5mm}Program~$P1$: Verification Conditions obtained by VCG}}
				\end{enumerate}	
			\end{multicols}
		}
		\vspace{-3mm}
	\end{minipage}
	\caption{Program~$P1$ is the set of Verification Conditions VCs 
		obtained by applying the VCG strategy starting from the C program~$P$, the initial condition ${\tt x}\geq0$ 
		and the error property ${\tt y}\leq0$.
		\label{fromCtoVCs}}
	\vspace{-2mm}
\end{figure}

%\medskip
In particular, NLR replaces the predicates \texttt{newp1}
and \texttt{newp2}, which are called with the non-linking variables \texttt{X2}, \texttt{Y1}, and \texttt{Y2}
(see clauses \ref{auxex4} and \ref{auxex3} of $P1$ in Figure~\ref{fromCtoVCs}), with two new predicates \texttt{newp3} and \texttt{newp4}, 
respectively, whose arguments are linking variables only.
Note that the removal of the two arguments {\tt Y1} and~{\tt X2} of {\tt newp1},
which are the non-linking variables in clause~\ref{auxex4}, determines in clause~\ref{auxex3} 
the removal of the two arguments {\tt Y1} and~{\tt X2}, which are  {\it linking} variables 
of {\tt newp2}. Thus, from {\tt newp2} with six arguments in clause~\ref{auxex3}, 
by removing 
also the non-linking variable~{\tt Y2}, 
we get the predicate {\tt newp4} in 
clauses~3' and~4'  
of program $P2$ with three arguments only (see Figure \ref{VCs-by-NLR-and-cFAR}).

\smallskip
The NLR strategy
consists in a repeated application of the \textit{unfolding},  \textit{definition introduction}, and \textit{folding} transformation rules~\cite{EtG96}.

We assume that the input of NLR is any CLP program {\it Prog}.
To keep the notation simple, we will identify a tuple of variables with the set
of variables occurring in it. The union of two tuples is constructed by erasing
duplicate elements.

During the execution the NLR strategy maintains in a set \textit{Defs}  all 
the definitions that have been introduced so far.
Every definition clause in \textit{Defs} is unfolded with respect 
to the leftmost atom in its body,
thereby producing a set ~{\textit{S}} of clauses.
Then every clause in \textit{S} is folded (repeatedly, with respect all atoms in its body) by using either definitions that already occur in \textit{Defs} or new definitions
that are introduced in \textit{Defs} for performing those folding steps.

The peculiarity of the NLR strategy lies in the 
careful management of the set of variables occurring 
in the head of the definition clauses.

Let  $C$ be a clause in $\textit{S}$ of the form: ~$\texttt{H\,:-\,c,\,L,\,B,\,R}$,
where the predicate symbol of \texttt{B} occurs in {\it Prog}.
If  $C$ cannot be folded with respect to~the atom \texttt{B} using any clause in {\it Defs}, 
then we have to introduce a new definition clause as we now explain.

First, we consider a definition $F$ whose head contains only the linking variables of the atom \texttt{B} in the clause $C$.
Let $F$ be $\texttt{newp(V):-\,B}$, where \texttt{newp} is a predicate symbol not occurring in the set ${\it Prog}\cup \textit{Defs}$, and $\texttt{V}$ is the set 
$\textit{linkvars}(\texttt{B},C)$ of the linking variables of \texttt{B} in \textit{C}.

{If} the set $\textit{Defs}$ contains a clause $D$ of the form $\texttt{newq(Q)\,:-\;S}$
such that, for some renaming substitution~$\vartheta$, $\texttt{B}\vartheta = \texttt{S}$,
{then} 
we replace clause $D$ in $\textit{Defs}$ with the clause 
$\texttt{newp(L):-\,B}$, where $\texttt{L}\!=\!\texttt{V}\vartheta\cup \texttt{Q}$.
{Otherwise}, we introduce the definition clause $F$ and we add it to  $\textit{Defs}$.

\smallskip
The introduction of the definition $F$ might seem to be the best choice
in the sense that it contains exactly the head variables which are actually needed  
for folding  clause~$C$.
However,  (variants of) \texttt{B} may occur also
in some other clauses to be folded.
Thus, if we directly introduce definitions whose heads contain linking variables only,
we run the risk of generating several definitions with the same atom in the body
and different sets of variables in the head (modulo renaming).

In order to keep the number of definitions low (and this will often improve the ability
of proving program correctness), 
instead of introducing multiple 
definitions containing the same atom in the body,
by applying the NLR strategy, we merge them in a single definition
whose set of head variables is the union 
of the head variables occurring in the merged definitions (modulo renaming).

The NLR strategy terminates when all clauses in $\textit{Defs}$\/  have been unfolded 
and no new definition need to be introduced for folding.

\begin{theorem}[Termination and Correctness of the NLR Strategy]
\label{thm:NLR}
Given any CLP program {\it Prog}, the NLR strategy terminates and produces 
a CLP program ${\it Prog}'$ such that {\tt unsafe} $\in \mathcal M({\it Prog})$ holds iff\/ 
{\tt unsafe} $\in \mathcal M({\it Prog}')$ holds.	
\end{theorem}

\paragraph{2. Constrained FAR Algorithm (cFAR){\rm{.}}}

Now we present an extension to constraint logic programs of the FAR algorithm 
presented in~\cite{LeS96} for removing redundant arguments from logic programs.
This extension will be called constrained FAR algorithm, or cFAR, for short. 
The objective of the FAR algorithm is to remove arguments that  
are not actually used during any computation of the program at hand. 
Indeed, it has been shown in~\cite{HeG06} that the FAR algorithm 
(and thus, also the cFAR algorithm) can be seen as 
a generalization of the liveness analysis.

\smallskip
In Figure~\ref{VCs-by-NLR-and-cFAR} we show the effect of applying the cFAR algorithm 
to the CLP program $P2$ obtained by the NLR strategy.
The output of the algorithm is the CLP program $P3$. Note that 
in program~$P3$ the predicate 
symbol {\tt newp4} denotes a different relation with respect to 
the one in program~$P2$, because in~$P3$ it has arity~2 and not~3.

\begin{figure}
{\small
\begin{multicols}{2}
\begin{enumerate}[topsep=-1pt,parsep=-1pt,partopsep=2pt,
	itemsep=1pt,labelsep*=5pt,leftmargin=18pt,label=\arabic*'.]	
	\item\label{auxex1p}{\tt	unsafe:- X1>=0, Y2=<0, newp3(X1,Y2).  }
	\item\label{auxex2p}{\tt	newp3(X1,Z2):- Z1=X1+1, newp4(X1,Z1,Z2).}	
	\item\label{auxex3p}{\tt	newp4(\underline{X1},Z1,Z2):- Z1=<9,~Z3=Z1+1,}	
	\item[]{\tt	~~~~~~~newp4(\underline{X1},Z3,Z2).	       }
	\item\label{auxex4p}{\tt	newp4(\underline{X1},Z1,Z1):- Z1>=10. }	
	\vspace{3mm}
	\item[]$P2$: Verification Conditions obtained by NLR
\end{enumerate}	
	\columnbreak
	\begin{enumerate}[parsep=-1pt,partopsep=-2pt,
		itemsep=1pt,labelsep*=6pt,leftmargin=26pt,label=\arabic*''.]
		\item\label{auxex17}{\tt unsafe:- X1>=0, Y2=<0, newp3(X1,Y2). }
		\item\label{auxex16}{\tt newp3(X1,Z2):- Z1=X1+1, newp4(Z1,Z2). }	
		\item\label{auxex15}{\tt newp4(Z1,Z2):- Z1=<9, Z3=Z1+1,} 	
		\item[]{\tt	~~~~newp4(Z3,Z2). }
		\item\label{auxex14}{\tt newp4(Z1,Z1):- Z1>=10. }
		\vspace{3mm}
		\item[]$P3$: Verification Conditions obtained by  cFAR
	\end{enumerate}	
\end{multicols}
}
\vspace*{-5mm}
\caption{Program $P2$ and Program~$P3$ are the Verification Conditions VCs 
obtained by applying the NLR strategy and  the cFAR algorithm, respectively.\label{VCs-by-NLR-and-cFAR}{}}
\vspace{-2mm}
\end{figure}

\smallskip
In order to define the constrained FAR algorithm we need to introduce some preliminary notions, some of which have been adapted from~\cite{LeS96}.

\begin{definition}[Erasure, Erased Atom, Erased Clause, Erased Program]
{\rm(i)}~An {\em{erasure}} is  a set of pairs each of which is of the form~$(\mathtt{p},k)$, where $\mathtt{p}$ is a predicate symbol of arity $n$ and $1\!\leq\! k \! \leq\! n$.

\noindent
{\rm(ii)}~Given an erasure $E$ and an atom $\mathtt{A}$ whose predicate symbol is $\mathtt{p}$,
the {\em{erased atom}} $\mathtt{A}|_E$ is obtained by
dropping all the arguments that occur at position $k$, 
for some $(\mathtt{p},k) \in E$.

\noindent
{\rm(iii)}~Given an erasure $E$ and a clause $C$ {\rm (}respectively, a 
CLP program ${\it Prog}${\rm )}, the {\em{erased clause}} $C|_E$ {\rm (}respectively, 
the {\em{erased program}} ${\it Prog}|_E${\rm )} is obtained by replacing all 
atoms $\mathtt{A}$ in $C$ {\rm (}respectively, in ${\it Prog}${\rm )} by $\mathtt{A}|_E$.
\end{definition}
In order to avoid the risk of collisions between predicate symbols after 
erasing some arguments, we assume 
that ${\it Prog}$ does not contain identical predicate symbols with different arity.

\smallskip
Obviously, we are interested in removing redundant arguments 
without altering the semantics of the original program, in the sense captured by the following definition.

\begin{definition}[Correctness of Erasure]
An erasure $E$ is \textit{correct} for a program ${\it Prog}$ if, for all atoms $\mathtt{A}$, we have that\,{\rm :} ~$\mathtt{A} \in \mathcal M({\it Prog})$ ~iff~ $\mathtt{A}|_E \in \mathcal M({\it Prog}|_E)$.
\end{definition}

Since we are dealing with constraint logic programs, the notion of multiple occurrences of a variable which is used in the original formulation of FAR~\cite{LeS96}, needs to be generalized as follows.
In this paper we assume that a constraint is a conjunction of $h\ (\geq 0)$
atomic constraints in the theory $\mathcal A$ of the linear integer arithmetics with integer arrays.

\begin{definition}[Variable Constrained to Another Variable]
Given two variables $\mathtt{X}$ and  $\mathtt{Y}$
and a constraint $\mathtt{c}$ of the form $\mathtt{c_1}\!\wedge \ldots \wedge\!\mathtt{c_h}$,
we say that   
$\mathtt{X}$ is \emph{constrained to}  $\mathtt{Y}$ {\rm (}in $\mathtt{c}${\rm )}
if there exists  $\mathtt{c_j}$, with $1\!\leq\!j\!\leq h$,
such that either {\rm(i)}  $\{\mathtt{X, Y}\} \subseteq \textit{vars}(\mathtt{c_j})$,
or {\rm(ii)}  there exists a variable $\mathtt{Z}$ such that {\rm(ii.1)} 
$\{\mathtt{X, Z}\} \subseteq \textit{vars}(\mathtt{c_j})$ and {\rm(ii.2)} $\mathtt{Z}$
is {constrained to}  $\mathtt{Y}$ {\rm (}in $\mathtt{c}${\rm )}.
\end{definition}

\smallskip
Now we are ready to introduce the notion of \textit{safe erasure} that will be used during the application of the constrained FAR algorithm.

\begin{definition}[Safe Erasure]
\label{def:safe_erasure}
Given a program ${\it Prog}$, an erasure $E$ is a {\em safe erasure} if, for all 
$(\mathtt{p},k) \in E$ and clauses $\mathtt{H:-c,G}$ in ${\it Prog}$,
where $\texttt{H}$ is of the form $\mathtt{p(X1,...,Xn)}$ 
and  $\mathtt{c}$ is of the form $\mathtt{c_1}\!\wedge \ldots \wedge\!\mathtt{c_h}$,
we have that:
{\rm(i)}~$\mathtt{X_k}$ is a variable
and { $\mathcal{A}\models \forall \mathtt{X_k.} \exists {\tt Y1,\ldots,Ym.}\ {\tt c}$,}
with  $\{{\tt Y1,\ldots,Ym}\} = {\textit{vars}({\tt c})-\{\mathtt{X_k}\}}$,
{\rm(ii)}~$\mathtt{X_k}$ is not constrained to any other variable occurring in $\mathtt{H}$, and
{\rm(iii)}~$\mathtt{X_k}$ is not constrained to any variable occurring in $\mathtt{G}|_E$.
\end{definition}

Similarly to what has been done in~\cite{LeS96}, it can be shown that if an erasure $E$ is {safe}, then it is also {correct}.

\smallskip
The cFAR algorithm takes as input a CLP program ${\it Prog}$, computes a safe erasure $E$, and produces as output the program ${\it Prog}|_E$.
The algorithm starts off by initializing the current erasure $E$ to  the  \textit{full erasure}, that is, the set of all pairs $(\texttt{p},k)$, where $p$ is a predicate of arity $n$ occurring in ${\it Prog}$ and $1\!\leq\! k \!\leq\! n$. 
Then, while $E$ contains a pair $(\texttt{p},k)$ 
such that one of the conditions of Definition~\ref{def:safe_erasure} is not satisfied, 
%for some clause in {\it Prog},
the pair~$(\texttt{p},k)$  is removed from $E$.
The algorithm terminates when it is no longer possible to remove 
a pair~$(\texttt{p},k)$ from $E$, and
 thus $E$ is a safe erasure.

\smallskip
The cFAR algorithm terminates and preserves the least-model semantics, as stated by the following theorem.

\begin{theorem}[Termination and Correctness of the cFAR Algorithm]\label{thm:FAR}
Given any CLP program ${\it Prog}$, the cFAR algorithm terminates and produces 
a CLP program ${\it Prog}|_E$ such that ${\tt unsafe} \in \mathcal M({\it Prog})$ iff\/ ${\tt unsafe} \in \mathcal M({\it Prog}|_E)$.	
\end{theorem}

Finally, we would like to note that, even if the objectives of the NLR and cFAR transformations 
are similar, they work in a different way. 
While cFAR is goal independent,
NLR starts from the predicate {\tt unsafe}
and proceeds by unfolding in a goal directed fashion, similarly to \textit{redundant argument filtering}~\cite{LeS96}.
It can be shown that, in general,  the NLR and cFAR transformations  have incomparable effects.

\section{Experimental evaluation}

We have used the VeriMAP transformation and verification 
system{~\cite{De&14c,De&14b}} 
for evaluating the techniques presented in this paper.
We have considered 
320 verification problems for C programs
(227~of which were safe and the remaining~93 were unsafe).
We have applied the {\rm VCG} strategy
for generating the Verification Conditions VCs 
using a multi-step semantics \cite{De&15b}.
The C~programs
and the VCs we have generated are\,available 
at: {\tt{http:\!/\!/map.uniroma2.it/vcgen}}. 
Then, we have checked the satisfiability of the VCs  
by giving them as input to the Z3 Horn solver~\cite{DeB08} {using default options\footnote{Note that Z3, by default, runs the \textit{slice} transformation for reducing the number of variables in the signature of a predicate.}
and the PDR engine.}
Finally, we have applied the NLR and cFAR transformations presented 
in Section~\ref{sec:auxtransf} to evaluate the effect
of these transformations in terms of efficiency and efficacy
in the program verification tasks considered.

\begin{table*}[htbp]
	\begin{center}
		{\small
			\begin{tabular}{|l|l|r|r|r|}
				\hline
				&                   &  VCG ; Z3  & VCG ; NLR ; Z3    & VCG ; NLR ; cFAR ; Z3    \\\hline
				\textit{c}  & Correct answers   &  196    &    7    &   9    \\\hline
				\textit{s}  & ~~~~- safe problems   &  144    &    3    &   7    \\\hline
				\textit{u}  & ~~~~- unsafe problems &   52    &    4    &   2    \\\hline
				\textit{to} & Timeouts          &  124    &  117    & 108    \\\hline
				\textit{n}  & Total problems    &  320    &  124    & 117    \\\hline\hline
				$\textit{t}_{{\textrm{\,VCG}}}$ & VCG time          &   40.65 &   20.48 &   4.57 \\\hline
				$\textit{t}_{{\textrm{\,NLR}}}$ & NLR time          &   --    &   58.39 &   9.53 \\\hline
				$\textit{t}_{{\textrm{\,cFAR}}}$ & cFAR time          &   --    &   --    & 304.84 \\\hline
				% source: 7RAF_9FAR_Table3_stats.ods
				\textit{st} & Z3 solving time & 2704.95 &  988.15 & 649.56 \\\hline\hline
				\textit{tt} & Total time        & 2745.60 & 1067.02 & 968.50 \\\hline
				\textit{at} & Average time      &   14.01 &  152.43 & 107.61 \\
				\hline
			\end{tabular}
		}
		\caption{Verification results obtained by using Z3 on the output 
			generated by applying VCG and the auxiliary transformations NLR and cFAR. 
			The timeout limit time is 300 seconds. Times are in seconds.}
		\label{table:auxres}
	\end{center}
\end{table*}

\paragraph{Improving effectiveness of solving{\rm{.}}}
In Table~\ref{table:auxres} we show the experimental results obtained by using 
VeriMAP and Z3.
Column `VCG ; Z3' reports the results obtained by applying the VCG strategy and then the Z3 solver.
Column `VCG ; NLR ; Z3' reports the results obtained, for the problems {\it not solved} by `VCG ; Z3',
by applying VCG, followed by the NLR transformation, and then Z3.
Column `VCG~; NLR~; cFAR~; Z3' reports the results obtained, for the problems  {\it not solved} by `VCG ; NLR ; Z3',
by applying VCG, followed by NLR, then cFAR, and finally Z3.
Lines~(${\textit{t}}_{{\textrm{\,VCG}}}$), ($\textit{t}_{{\textrm{\,NLR}}}$), and
($\textit{t}_{{\textrm{\,cFAR}}}$) report the time taken by the execution of the 
{\rm VCG}, {\rm NLR}, and {\rm cFAR} transformations, respectively, to produce 
the verification conditions 
for which  Z3 was able to return the correct answers (that is, to show
the satisfiability or the unsatisfiability of the clauses).
Line ({\it st}) reports the time taken by Z3 to produce the correct answers.

The NLR transformation enables Z3 to prove 7 additional verification problems.
In particular, it allows Z3 to prove the program \texttt{ntdrvsimpl-cdaudio\_simpl1\_unsafeil.c}, 
which is the largest program in the benchmark set (2.1 KLOC). 
Concerning the time required for executing the NLR transformation in this
example, we want to point out
that this program takes 91\% of the total NLR time ($\textit{t}_{{\textrm{\,NLR}}}$), 
that is 53.04 seconds. Therefore, the remaining 6 programs only require 5.35 
seconds to be transformed.
The cFAR transformation allows Z3 to prove 9 additional verification problems.
In this case, about 89\% of the total cFAR time ($\textit{t}_{{\textrm{\,cFAR}}}$), 
that is, 271.62 seconds, is required for specializing two programs
whose size is about 1 KLOC each,
namely \texttt{ntdrvsimpl-diskperf\_simpl1\_safeil.c} (98.82 seconds) and
\texttt{ntdrvsimpl-floppy\_simpl3\_safeil.c} (172.80 seconds).

\section{Conclusions}
In this paper we have shown that the effectiveness of Horn clause solvers
for proving the satisfiability of VCs
can be improved by  the use of
program transformations
that remove unnecessary variables.

As future work, we would like to investigate in more depth 
how the structure of the VCs influences 
the heuristics adopted by Horn solvers.
Hopefully, this would allow us to tune the VCG strategy 
for generating VCs that are easier to be proved.
Also, it would be interesting to study the effect 
of the NLR and cFAR transformations on the VCs generated by other tools like, for example, SeaHorn~\cite{Gu&15}.

{\section*{Acknowledgements}
The authors would like to thank the GNCS - INdAM for the Research Grant 2016 ``Verifica Automatica di Propriet\`a Relazionali di Programmi''.}

\vspace{2mm}
%\nocite{*}
%\newpage 

\bibliographystyle{eptcs}

\bibliography{Smc,Transformation}

\begin{thebibliography}{10}
\providecommand{\bibitemdeclare}[2]{}
\providecommand{\surnamestart}{}
\providecommand{\surnameend}{}
\providecommand{\urlprefix}{Available at }
\providecommand{\url}[1]{\texttt{#1}}
\providecommand{\href}[2]{\texttt{#2}}
\providecommand{\urlalt}[2]{\href{#1}{#2}}
\providecommand{\doi}[1]{doi:\urlalt{http://dx.doi.org/#1}{#1}}
\providecommand{\bibinfo}[2]{#2}

\bibitemdeclare{incollection}{Al&07}
\bibitem{Al&07}
\bibinfo{author}{E.~\surnamestart Albert\surnameend},
  \bibinfo{author}{M.~\surnamestart G\'{o}mez-Zamalloa\surnameend},
  \bibinfo{author}{L.~\surnamestart Hubert\surnameend} \&
  \bibinfo{author}{G.~\surnamestart Puebla\surnameend} (\bibinfo{year}{2007}):
  \emph{\bibinfo{title}{Verification of {J}ava {B}ytecode {U}sing {A}nalysis
  and {T}ransformation of {L}ogic {P}rograms}}.
\newblock In \bibinfo{editor}{M.~\surnamestart Hanus\surnameend}, editor: {\sl
  \bibinfo{booktitle}{Practical {A}spects of {D}eclarative {L}anguages}},
  \bibinfo{series}{Lecture Notes in Computer Science 4354},
  \bibinfo{publisher}{Springer}, pp. \bibinfo{pages}{124--139},
  \doi{10.1007/978-3-540-69611-7\_8}.

\bibitemdeclare{article}{De&14c}
\bibitem{De&14c}
\bibinfo{author}{E.~\surnamestart {De~Angelis}\surnameend},
  \bibinfo{author}{F.~\surnamestart Fioravanti\surnameend},
  \bibinfo{author}{A.~\surnamestart Pettorossi\surnameend} \&
  \bibinfo{author}{M.~\surnamestart Proietti\surnameend}
  (\bibinfo{year}{2014}): \emph{\bibinfo{title}{Program Verification via
  Iterated Specialization}}.
\newblock {\sl \bibinfo{journal}{Science of Computer Programming}}
  \bibinfo{volume}{95, Part 2}, pp. \bibinfo{pages}{149--175},
  \doi{10.1016/j.scico.2014.05.017}.

\bibitemdeclare{inproceedings}{De&14b}
\bibitem{De&14b}
\bibinfo{author}{E.~\surnamestart {De~Angelis}\surnameend},
  \bibinfo{author}{F.~\surnamestart Fioravanti\surnameend},
  \bibinfo{author}{A.~\surnamestart Pettorossi\surnameend} \&
  \bibinfo{author}{M.~\surnamestart Proietti\surnameend}
  (\bibinfo{year}{2014}): \emph{\bibinfo{title}{{V}eri{MAP}: {A} {T}ool for
  {V}erifying {P}rograms through {T}ransformations}}.
\newblock In: {\sl \bibinfo{booktitle}{Proceedings of the 20th International
  Conference on Tools and Algorithms for the Construction and Analysis of
  Systems, TACAS~'14}}, \bibinfo{series}{Lecture Notes in Computer Science
  8413}, \bibinfo{publisher}{Springer}, pp. \bibinfo{pages}{568--574},
  \doi{10.1007/978-3-642-54862-8\_47}.
\newblock \bibinfo{note}{Available at: {\rm
  http://www.map.uniroma2.it/VeriMAP}}.

\bibitemdeclare{inproceedings}{De&15b}
\bibitem{De&15b}
\bibinfo{author}{E.~\surnamestart {De Angelis}\surnameend},
  \bibinfo{author}{F.~\surnamestart Fioravanti\surnameend},
  \bibinfo{author}{A.~\surnamestart Pettorossi\surnameend} \&
  \bibinfo{author}{M.~\surnamestart Proietti\surnameend}
  (\bibinfo{year}{2015}): \emph{\bibinfo{title}{Semantics-based generation of
  verification conditions by program specialization}}.
\newblock In: {\sl \bibinfo{booktitle}{Proceedings of the 17th International
  Symposium on Principles and Practice of Declarative Programming, Siena,
  Italy, July 14-16, 2015}}, \bibinfo{publisher}{ACM}, pp.
  \bibinfo{pages}{91--102}, \doi{10.1145/2790449.2790529}.

\bibitemdeclare{article}{EtG96}
\bibitem{EtG96}
\bibinfo{author}{S.~\surnamestart Etalle\surnameend} \&
  \bibinfo{author}{M.~\surnamestart Gabbrielli\surnameend}
  (\bibinfo{year}{1996}): \emph{\bibinfo{title}{Trans\-form\-ations of {CLP}
  Modules}}.
\newblock {\sl \bibinfo{journal}{Theoretical Computer Science}}
  \bibinfo{volume}{166}, pp. \bibinfo{pages}{101--146},
  \doi{10.1016/0304-3975(95)00148-4}.

\bibitemdeclare{article}{GaK14a}
\bibitem{GaK14a}
\bibinfo{author}{J.~P. \surnamestart Gallagher\surnameend} \&
  \bibinfo{author}{B.~\surnamestart Kafle\surnameend} (\bibinfo{year}{2014}):
  \emph{\bibinfo{title}{Analysis and {T}ransformation {T}ools for {C}onstrained
  {H}orn {C}lause {V}erification}}.
\newblock {\sl \bibinfo{journal}{Theory and Practice of Logic Programming}}
  \bibinfo{volume}{14}(\bibinfo{number}{4-5}), pp. \bibinfo{pages}{90--101}.
\newblock \bibinfo{note}{Supplementary Materials}.

\bibitemdeclare{inproceedings}{Gu&15}
\bibitem{Gu&15}
\bibinfo{author}{A.~\surnamestart Gurfinkel\surnameend},
  \bibinfo{author}{T.~\surnamestart Kahsai\surnameend},
  \bibinfo{author}{A.~\surnamestart Komuravelli\surnameend} \&
  \bibinfo{author}{J.A. \surnamestart Navas\surnameend} (\bibinfo{year}{2015}):
  \emph{\bibinfo{title}{The {SeaHorn} {V}erification {F}ramework}}.
\newblock In: {\sl \bibinfo{booktitle}{Computer Aided Verification: 27th
  International Conference, CAV 2015, San Francisco, CA, USA, July 18-24,
  2015}}, \bibinfo{publisher}{Springer}, pp. \bibinfo{pages}{343--361},
  \doi{10.1007/978-3-319-21690-4\_20}.

\bibitemdeclare{inproceedings}{HeG06}
\bibitem{HeG06}
\bibinfo{author}{K.~S. \surnamestart Henriksen\surnameend} \&
  \bibinfo{author}{J.~P. \surnamestart Gallagher\surnameend}
  (\bibinfo{year}{2006}): \emph{\bibinfo{title}{Abstract Interpretation of PIC
  Programs through Logic Programming}}.
\newblock In: {\sl \bibinfo{booktitle}{Proceedings of the 6th IEEE
  International Workshop on Source Code Analysis and Manipulation, SCAM~'06}},
  pp. \bibinfo{pages}{103--179}, \doi{10.1109/SCAM.2006.1}.

\bibitemdeclare{inproceedings}{LeS96}
\bibitem{LeS96}
\bibinfo{author}{M.~\surnamestart Leuschel\surnameend} \&
  \bibinfo{author}{M.~H. \surnamestart S{\o}rensen\surnameend}
  (\bibinfo{year}{1996}): \emph{\bibinfo{title}{Redundant Argument Filtering of
  Logic Programs}}.
\newblock In \bibinfo{editor}{J.~\surnamestart Gallagher\surnameend}, editor:
  {\sl \bibinfo{booktitle}{Logic Pro\-gram Synthesis and Trans\-form\-ation,
  Proceedings LOPSTR '96, Stockholm, Sweden}}, \bibinfo{series}{Lecture Notes
  in Computer Science 1207}, \bibinfo{publisher}{Springer-Verlag}, pp.
  \bibinfo{pages}{83--103}, \doi{10.1007/3-540-62718-9\_6}.

\bibitemdeclare{inproceedings}{DeB08}
\bibitem{DeB08}
\bibinfo{author}{L.~M. \surnamestart de~Moura\surnameend} \&
  \bibinfo{author}{N.~\surnamestart Bj{\o}rner\surnameend}
  (\bibinfo{year}{2008}): \emph{\bibinfo{title}{Z3: {A}n Efficient {SMT}
  Solver}}.
\newblock In: {\sl \bibinfo{booktitle}{Proceedings of the 14th International
  Conference on Tools and Algorithms for the Construction and Analysis of
  Systems, {TACAS}~'08}}, \bibinfo{series}{Lecture Notes in Computer Science
  4963}, \bibinfo{publisher}{Springer}, pp. \bibinfo{pages}{337--340},
  \doi{10.1007/978-3-540-78800-3\_24}.

\bibitemdeclare{inproceedings}{Pe&98}
\bibitem{Pe&98}
\bibinfo{author}{J.~C. \surnamestart Peralta\surnameend},
  \bibinfo{author}{J.~P. \surnamestart Gallagher\surnameend} \&
  \bibinfo{author}{H.~\surnamestart Saglam\surnameend} (\bibinfo{year}{1998}):
  \emph{\bibinfo{title}{Analysis of {I}mperative {P}rograms through {A}nalysis
  of {C}onstraint {L}ogic {P}rograms}}.
\newblock In \bibinfo{editor}{G.~\surnamestart {L}evi\surnameend}, editor: {\sl
  \bibinfo{booktitle}{Proceedings of the 5th {I}nternational {S}ymposium on
  {S}tatic {A}nalysis, {SAS}~'98}}, \bibinfo{series}{Lecture Notes in Computer
  Science 1503}, \bibinfo{publisher}{Springer}, pp. \bibinfo{pages}{246--261},
  \doi{10.1007/3-540-49727-7\_15}.

\bibitemdeclare{article}{PrP95a}
\bibitem{PrP95a}
\bibinfo{author}{M.~\surnamestart Proietti\surnameend} \&
  \bibinfo{author}{A.~\surnamestart Pettorossi\surnameend}
  (\bibinfo{year}{1995}): \emph{\bibinfo{title}{Unfolding-Definition-Folding,
  in this Order, for Avoiding Unnecessary Variables in Logic Pro\-grams}}.
\newblock {\sl \bibinfo{journal}{Theoretical Computer Science}}
  \bibinfo{volume}{142}(\bibinfo{number}{1}), pp. \bibinfo{pages}{89--124},
  \doi{10.1016/0304-3975(94)00227-A}.

\end{thebibliography}

\end{document}